\documentclass{article}

\usepackage{PRIMEarxiv}

\usepackage[utf8]{inputenc} 
\usepackage[T1]{fontenc}    
\usepackage{hyperref}       
\usepackage{url}            
\usepackage{booktabs}       
\usepackage{amsfonts}       
\usepackage{nicefrac}       
\usepackage{microtype}      
\usepackage{lipsum}
\usepackage{fancyhdr}       
\usepackage{graphicx}       
\usepackage{multirow}
\graphicspath{{media/}}     

\title{Towards Measuring and Quantifying the Comprehensibility of Process Models - The Process Model Comprehension Framework
}

\author{
  Michael Winter$^{1}$*, Rüdiger Pryss$^{2}$, Matthias Fink$^{3}$, Manfred Reichert$^{1}$\\
	$^{1}$ Institute of Databases and Information Systems, Ulm University, Ulm, Germany; \\ \{michael.winter,manfred.reichert\}@uni-ulm.de\\
	$^{2}$ Institute of Clinical Epidemiology and Biometry, University of W\"urzburg, W\"urzburg, Germany; \\ ruediger.pryss@uni-wuerzburg.de \\
	$^{3}$ Ventum Consulting GmbH \& Co. KG, Infanteriestra{\ss}e 11A, 80797 M\"unchen; \\ matthias.fink@ventum.de \\
	**Corresponding author\\
	\\
	Preprint puplished on arXiv.org
}

\begin{document}
\maketitle

\begin{abstract}
Process models constitute crucial artifacts in modern information systems and, hence, the proper comprehension of these models is of utmost importance in the utilization of such systems. Generally, process models are considered from two different perspectives: process modelers and readers. Both perspectives share similarities and differences in the comprehension of process models (e.g., diverse experiences when working with process models). The literature proposed many rules and guidelines to ensure a proper comprehension of process models for both perspectives. As a novel contribution in this context, this paper introduces the Process Model Comprehension Framework (PMCF) as a first step towards the measurement and quantification of the perspectives of process modelers and readers as well as the interaction of both regarding the comprehension of process models. Therefore, the PMCF describes an Evaluation Theory Tree based on the Communication Theory as well as the Conceptual Modeling Quality Framework and considers a total of 96 quality metrics in order to quantify process model comprehension. Furthermore, the PMCF was evaluated in a survey with 131 participants and has been implemented as well as applied successfully in a practical case study including 33 participants. To conclude, the PMCF allows for the identification of pitfalls and provides related information about how to assist process modelers as well as readers in order to foster and enable a proper comprehension of process models.
\end{abstract}


\section{Introduction}

Business Process Management (BPM) describes the discipline in bridging the gap between business, technology, and human workers in organizations \cite{rahimi2016business}. In more detail, modern information technologies (e.g., process-aware information systems; PAIS) are the enabler towards the automation of the processes in organizations and comprising the interaction between humans and the application of technology (i.e., human-driven processes) \cite{dumas2005process}. As a prerequisite for the successful utilization of PAIS, it must be ensured that the numerous processes of organizations are comprehended correctly and that respective information as well as knowledge about these processes are proper documented; either textually or visually \cite{zimoch2017using}. Thereby, in order to sustain competitive advantage, an easy to comprehend and correct documentation of process information and knowledge is essential \cite{polyvyanyy2015business}. In this context, an established approach to document process information and knowledge is to rely on the technique of process modeling, in which respective information as well as knowledge are visually documented in process models. More specifically, process models summarize the individual processes of organizations with their logical sequence of activities and functions, together with involved stakeholders or exchanged data. For this reason, one of the main purposes of process models is to communicate information and knowledge about corresponding processes. As a result, process models should be created in a way that involved stakeholders do not encounter any challenges in the comprehension of such models in order to take full advantage of their benefits \cite{awadid2019consistency}. 

In general, all stakeholders involved in working with process models can be assigned into a group consisting of process modelers, process readers, or a combination of both \cite{zimoch2017eye}. Initially, a process modeler consolidates required information and knowledge about a process and, hence, creates a corresponding process model based on it. Thereby, the process modeler should be aware that the created process model reflects a high model quality in order to ensure a proper process model comprehension \cite{de2015systematic}. Accordingly, with the assistance of the created process model, process readers are able to extract information as well as knowledge about related processes. However, a process model of high quality that is comprehensible for the initial model creator does not ensure that even a process reader is able to comprehend the same model \cite{MendlingRRL19}. Usually, the two major reasons for this are, on one the hand, that there exists a gap of experience and expectations (i.e., different perspectives) in working with process models \cite{figl2017comprehension}. On the other hand, pitfalls (e.g., modeling errors) in the communication of process knowledge as well as information between process modelers and readers describe another reason \cite{haisjackl2018humans}. To tackle these issues, specific guidelines and frameworks in literature exist putting an emphasis on quality aspects (e.g., consistency in process models) to foster the comprehension of conceptual as well as process models. For example, one of the most influential frameworks for conceptual modeling constitutes the SEQUAL framework introduced in \cite{lindland1994understanding}. This framework considers three different quality dimensions (i.e., syntactic, semantic, and pragmatic quality) and provides means of improvement for each quality dimensions in order to maintain a high quality in conceptual models, thus having a positive influence on the comprehension of such models. Further, the authors in \cite{krogstie2006process} are addressing shortcomings (e.g., static view upon semantic quality) of the SEQUAL framework and propose an adjusted framework. In addition, a significant enhancement of this work describes the consideration of as-is as well as to-be states (i.e., domain and knowledge). Based on semiotics, an integrative framework for information systems is discussed in \cite{mingers2014integrative}. Thereby, the authors consider the interaction of three worlds (i.e., material, personal, and social) derived from Sociomateriality Theory and use this kind of interaction to discuss deficits and improvements in model comprehension. Another framework for the evaluation of the quality and comprehension in conceptual models constitutes the Bung-Wand-Weber (BWW) framework \cite{gehlert2007bww}. It comprises metrics to evaluate the quality in conceptual models. Thereby, a focus is set on the modeling process and the BWW framework considers how objects from the real world change during the transformation into a conceptual model and the impact on the model quality as well as comprehension during this transformation. Moreover, the Guidelines of Modeling (GoM) describe another framework to measure the quality in process models from different viewpoints in order to foster model comprehension \cite{becker2000guidelines}. In this context, the work presented in \cite{mendling2010seven} describes a set of seven process modeling guidelines (7PMG) assisting process modelers in the creation of comprehensible models. Finally, the work presented in \cite{de2018overview} introduces the Comprehensive Process Model Quality Framework (CPMQF). The CPMQF summarizes existing knowledge about process model quality and structures related knowledge based on six key questions, with an emphasis on completeness and relevance of quality aspects in process models. However, all discussed works are mainly on a theoretical basis and none provides an applicable measurement and the quantification of the perspectives of process modelers and readers in process model comprehension. As a consequence, the identification of aspects in a process model that are hard to comprehend (i.e., noise) is still tedious, because the results presented in the discussed works might be too abstract (i.e., no clear directional guidance for process model improvement). In addition, especially novices or non-experts may find it difficult to recognize their benefits in the context of process model comprehension. 

For this reason, in line with prior conducted research and as a further contribution to improve our understanding of working with process models, we try to foster process model comprehension with an approach that recapitulates and quantifies the specific perspectives of process modelers and readers as well as the interaction between both groups as main determinants in model comprehension. Therefore, this paper presents the Process Model Comprehension Framework (PMCF). The PMCF describes the first step towards a framework to measure the comprehensibility of process models from the perspective of process modelers, readers, and the interaction of both. Therefore, in a consensus building process with experts from BPM and existing literature, an Evaluation Theory Tree (ETT) with 96 quality metrics was defined. The ETT was evaluated in a survey with 131 students and practitioners to determine the importance and degree of impact on process model comprehension of the quality criteria as well as metrics used in the ETT. In conclusion, the PMCF quantifies process model comprehension on evaluated process models taking both the perspective of process modelers and readers into account. To demonstrate the applicability of the PMCF, a case study with 33 participants from industry was conducted. In general, the PMCF shall unravel general pitfalls that needed to be addressed in order to ensure a proper comprehension of process models. Furthermore, a uniform model comprehensibility is pursued with the application of the PMCF between process modelers and readers. In the future, the PMCF is intended to provide additional assistance for organizations in the efficient and effective utilization of information systems.  

The structure of this paper is as follows: Section 2 provides theoretical fundamentals of the PMCF. The PMCF and the defined ETT are presented in Section 3. Section 4 describes the implementation of the PMCF. In Section 5, the PMCF is demonstrated in a case study. In addition, based on the case study, Section 5 presents how existing process models in a practical environment can be improved in terms of process model comprehension with the PMCF. Furthermore, current limitations as well as implications of the PMCF and future work are discussed in this section. Finally, Section 6 summarizes the paper.

\section{Theoretical Fundamentals}

This section introduces the underlying theoretical fundamentals of the PMCF: the Communication Theory (see Section 2.1) and the Conceptual Modeling Quality Framework (CMQF) (see Section 2.2). Figure 1 illustrates the theoretical fundamentals with their related contribution (i.e., green), their emerging issue (i.e., red), and corresponding focus (i.e., blue) of the subsequent fundament. 

\begin{figure}[h]
	\centering
	\includegraphics[width=\linewidth]{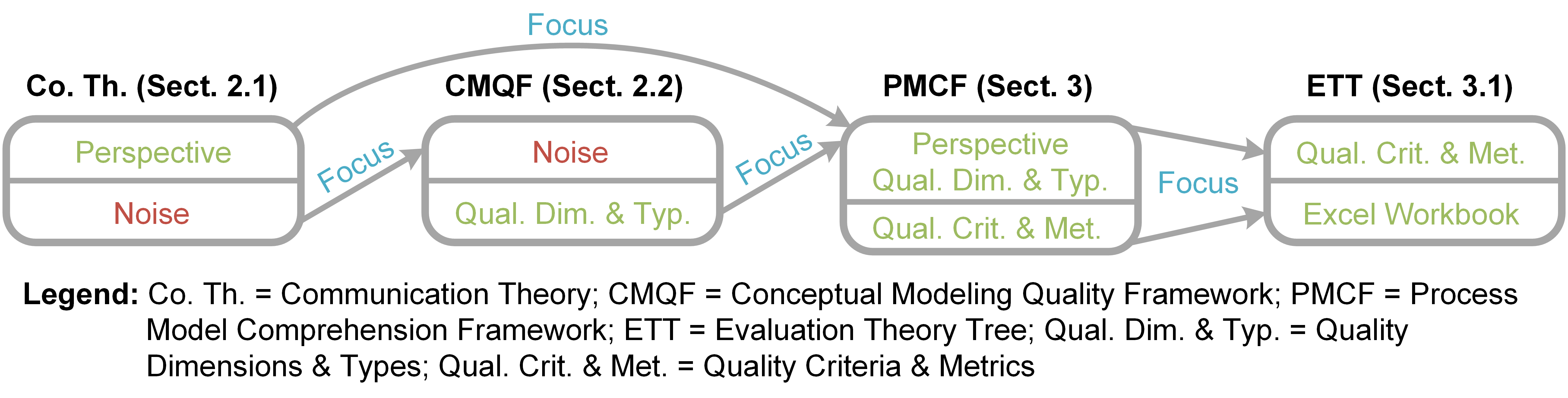}
	\caption{Theoretical Fundamentals and their focus}
\end{figure}

\subsection{Communication Theory}

According to the Communication Theory (see Figure 2), a process model constitutes an artifact utilized for the communication of information and knowledge about a process between two participants \cite{moody2009physics}. Thereby, the two participants involved in this kind of communication can be denoted as transmitter (i.e., process modeler) and receiver (i.e., process reader) of information and knowledge. More specifically, the process modeler encodes respective information and knowledge about a process within a medium. In this context, the medium describes a process model. In general, a process model delineates a conceptual model that is used to transfer information and knowledge about a subject the model represents (e.g., order to cash process) \cite{da2015model}. Thereby, a process model is expressed in terms of a particular process modeling language (e.g., Business Process Model and Notation (BPMN)), which is used to communicate information and knowledge about events, activities, decisions, data, and involved participants \cite{kocbek2015business}. Thereby, a process modeling language is described by two components: (1) alphabet (i.e., set of graphemes) and (2) grammar (i.e., systematic description of the process modeling language). Hence, it is important that a process modeler has an adequate understanding of the alphabet and the corresponding grammar for the proper documentation of process information and knowledge in a process model. In turn, captured information and knowledge in a process model is decoded by a process reader. In decoding, the human perception constitutes the central information processing system and describes the two psychological processes (a) visual perception (i.e., processing of visual information and knowledge) and (b) comprehension (i.e., interpretation of information and knowledge). As a consequence, the encoding as well as the decoding of information and knowledge in a process model results in different perspectives for process modelers and readers. Consequently, pitfalls (i.e., noise) may occur between the communication of process modelers and readers. In particularly, noise defines perturbations in the comprehension of process models. These perturbations cause ambiguities between process modelers as well as readers regarding the communicated information and knowledge in a process model, thus leading to a non-uniform process model comprehension. For example, the conception of the process modeler about the process or the used modeling language for the creation of a corresponding process model are potential noise factors in the encoding phase \cite{claes2017structured}. Regarding the process model, the intention (e.g., process optimization), with which the process model (e.g., textual or visual) is perceived, denotes another noise factor \cite{de2018origin}. Finally, reasons for noise in the decoding phase are mainly the perceptual as well as cognitive processing (e.g., expertise in working with process models) of information and knowledge in the process model \cite{caivano2018artifact}. Generally, the occurrence of noise in this context depends on many additional factors \cite{trkman2019impact,mendling2007understanding,figl2017comprehension}. 

A significant reason for the occurrence of noise between the three aspects encoding, process model, and decoding is mainly due to the lack of the overall process model quality in this communication procedure \cite{krogstie2016quality}. Thereby, quality defines characteristics aspects (e.g., process modeling expertise, correctness of a process model) that can be measured and compared with each other (e.g., degree of excellence) \cite{ghicajanu2015criteria}. In this context, the Conceptual Modeling Quality Framework (CMQF), therefore, defines a set of quality aspects in order to prevent noise and, at the same time, to assure a high quality in the creation and comprehension of conceptual models (e.g., process model). 

\begin{figure}[h]
	\centering
	\includegraphics[width=\linewidth]{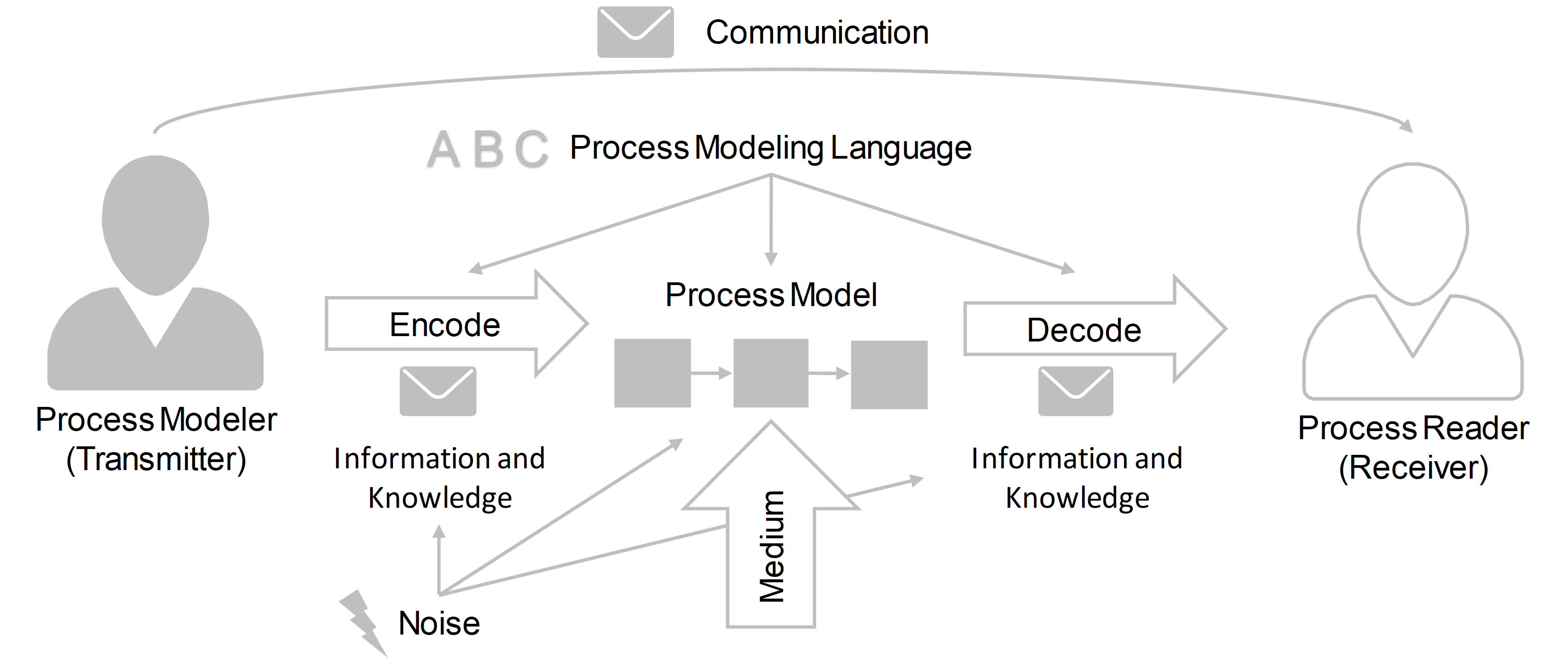}
	\caption{Communication Theory}
	\label{fig:ct}
\end{figure}

\subsection{Conceptual Modeling Quality Framework}

The Conceptual Modeling Quality Framework (CMQF) presents a unified overview considering the quality of the conceptual modeling process as well as the quality of the corresponding final result (i.e., conceptual model) \cite{nelson2012conceptual}. Figure 3 presents the CMQF with corresponding clusters, dimensions, and layers with related quality types (i.e., physical: red, knowledge: green, learning: purple, development: blue). In general, the CMQF addresses figuratively occurring noise known from the Communication Theory (see Section 2.1) without having a concrete perspective of a process modeler nor a reader. Importantly, the CMQF defines two horizontal clusters describing the physical (i.e., real world) and the cognitive reality (i.e., cognitive perception). The physical reality refers to the domain of discourse \cite{regoczei1987creating}, whereas the cognitive reality describes the constructed representation of the perception from the real world. Moreover, for each horizontal cluster, the CMQF defines four vertical clusters: domain, model, language, and representation. These four vertical clusters represent the conceptual modeling process. In particular, the domain refers to the process environment, which can be depicted in a conceptual model. The conceptual model, in turn, is created in terms of a particular modeling language resulting in a specific representation of the conceptual model. Moreover, all clusters comprise eight different quality dimensions. These eight quality dimensions constitute either physical or cognitive artifacts in conceptual modeling. Moreover, the quality dimensions are associated with quality types, summarized in four different layers: the physical (see Figure 3, red), knowledge (see Figure 3, green), learning (see Figure 3, purple), and development (see Figure 3, blue) layer. In the physical layer, the appropriateness of a conceptual model for the depiction of a process and its environment is evaluated. The knowledge layer states that for each physical representation a cognitive equivalent representation in the perception exists. Furthermore, the learning layer explains how information and knowledge are acquired by the interpretation of the real world. Finally, the development layer describes that knowledge and information are used in the creation of physical artifacts (e.g., conceptual model). The quality types define for each layer the relationship between a {reference} and a {purpose of application}. More specifically, the reference constitutes the chosen quality dimension, whereas the purpose of application depicts the quality dimension that is being considered across all quality dimensions. Moreover, to draw on the Communication Theory, the quality types are responsible for the prevention of noise. For example, the quality aspect between the physical domain (i.e., reference) and the domain knowledge (i.e., purpose of application) depends strongly on the perception of a person. Hence, it is of importance to ensure that the person has a correct understanding of the domain. The four layers as well as the quality types (i.e., seven in physical, seven in learning, four in learning, and six in development) depict the conceptual modeling process and, at the same time, preserve the completeness as well as the correctness of the final conceptual model.

\begin{figure}
	\centering
	\includegraphics[width=.7\linewidth]{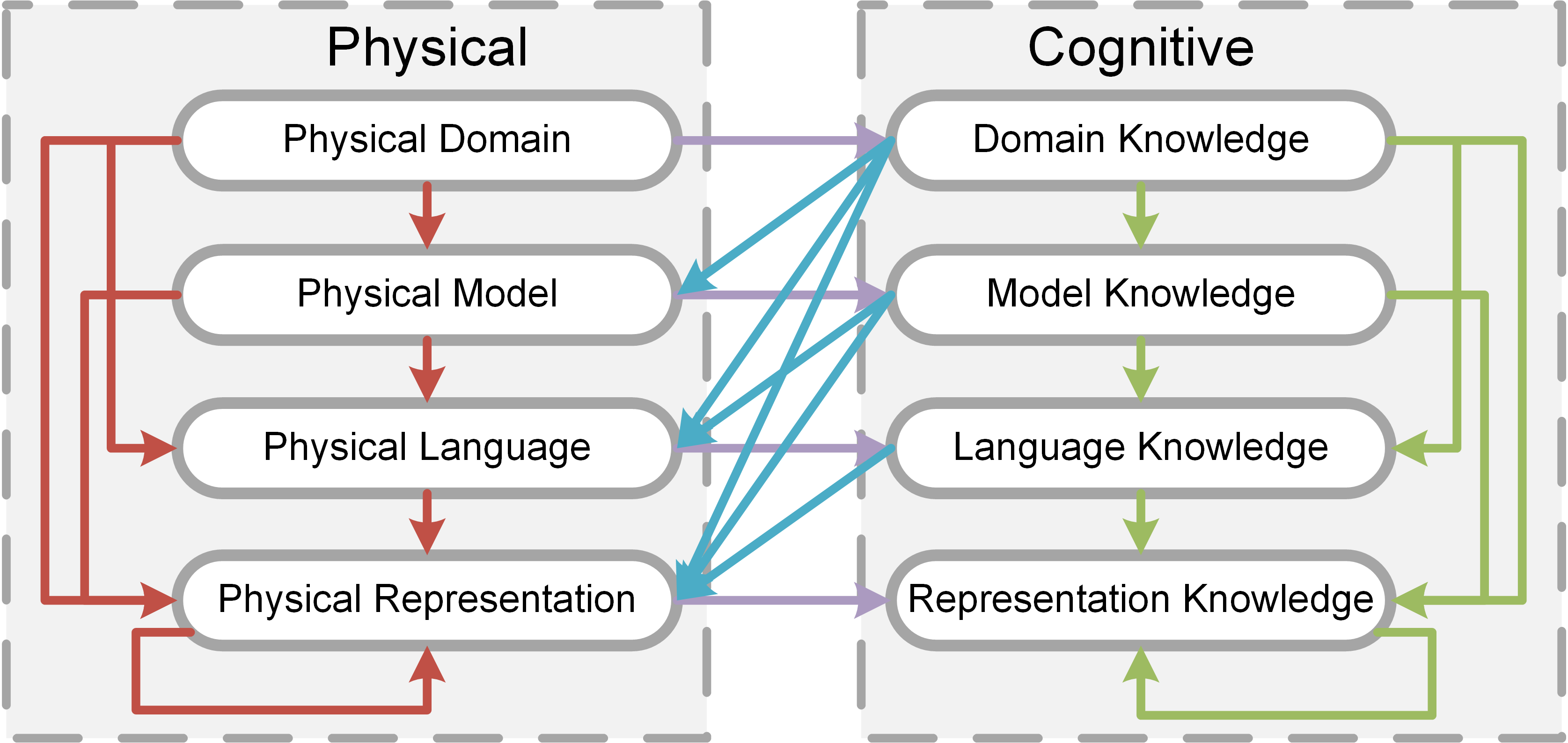}
	\caption{Conceptual Modeling Quality Framework (CMQF)}
	\label{fig:cmqf}
\end{figure}

\section{Process Model Comprehension Framework}

The Process Model Comprehension Framework (PMCF) is an adaption of the CMQF and considers the comprehension of process models based on the fundamentals of the Communication Theory (see Section 2). The PMCF allows for the measurement of the perspectives of process modelers and readers as well as the interaction between both as main determinants in the context of process model comprehension. As novelty, the PMCF is capable to quantify process model comprehension for different perspectives (i.e., process modelers and readers) and facilitates the identification of noise in model comprehension. Figure 4 delineates the PMCF. As known from the CMQF (see Fig. 3), the two vertical clusters (i.e., physical and cognitive reality), the four layers (physical (P1 - 7, red), knowledge (K1 - 7, green), learning (L1 - 4, purple), and development (D1 - 6, blue) remain unchanged in the PMCF. The four vertical clusters (i.e., domain, model, language, and representation), the inherent eight quality dimensions as well as the associated quality types have been adapted accordingly to fit to the requirements concerning process models. The first vertical cluster refers to the process and its environment as well as the process knowledge thereof. The second cluster, in turn, considers the process model and related model knowledge. Similarly as in the second cluster, the third cluster correlates the used process modeling language with respective knowledge. Finally, the fourth cluster describes the representation of the process in the real world and in the perception. In the PMCF, the same quality types from the CMQF are used, but are not considered as unidirectional relationships. Instead, the relationships between the quality types are bidirectional. The reason is that a quality type between two dimensions, on the one hand, leads to knowledge gain and, on the other hand, indicates the necessary knowledge level to assure a high model quality. For example, consider the quality type K1 in Fig. 4, this quality type between the Process Domain Knowledge (i.e., reference) and Process Model Knowledge (i.e., purpose of application) addresses the fact that describes which knowledge level about the process is required (e.g., information about value-adding activities) to represent this process in a process model. On the contrary, changing reference with purpose of application, describes that the comprehension of a process model consequently results in new insights (e.g., identification of bottlenecks) about the process. 

\begin{figure}[h]
	\centering
	\includegraphics[width=.9\linewidth]{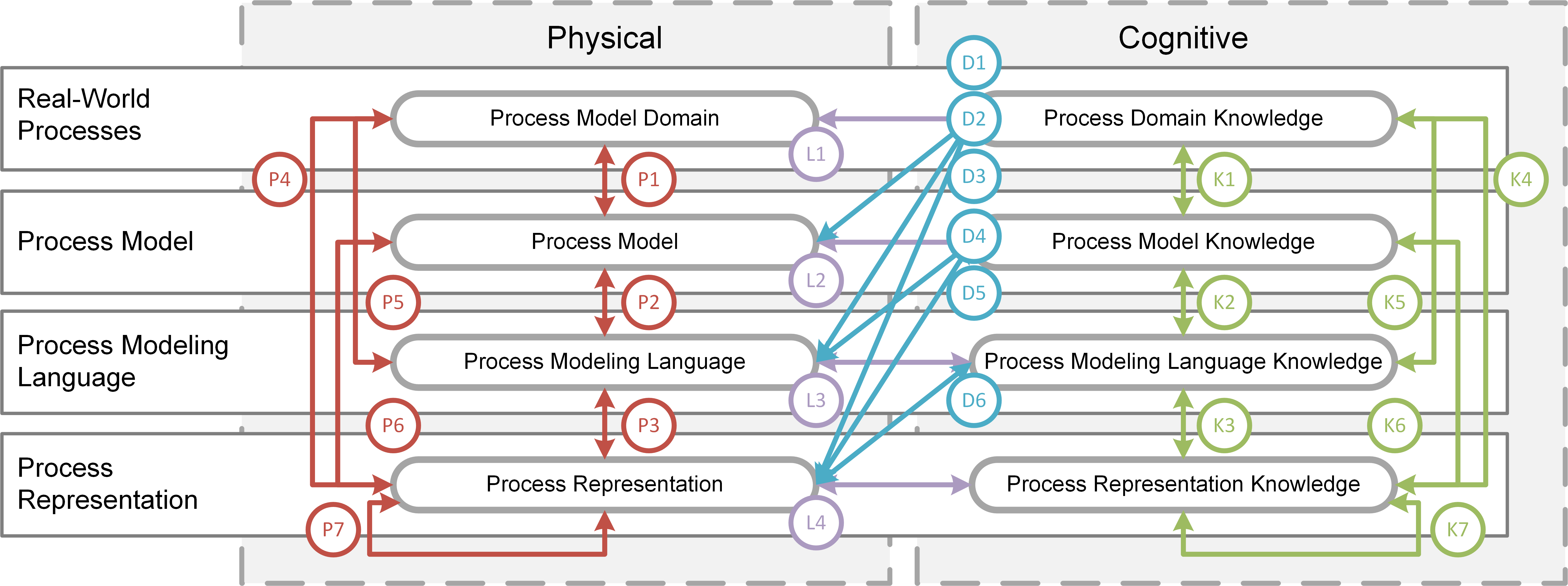}
	\caption{Process Model Comprehension Framework (PMCF)}
	\label{fig:pmfc}
\end{figure}

\subsection{Evaluation Theory Tree (ETT)}

Based on the fundamentals discussed, an Evaluation Theory Tree (ETT) was defined \cite{christie2008evaluation}. In general, the ETT represents a convolution of the PMCF. Hence, the roots of the ETT consider the perspectives of the process modelers and readers. The reason for two roots in the ETT is due to the fact that aspects exist that cannot be mapped directly between both perspectives. For example, the creation of a process model is only relevant for process modelers. Therefore, process modelers and readers must be considered separately. Each perspective, in turn, consists of a number of aggregated quality criteria in the context of process model comprehension. These quality criteria are related to the eight quality dimensions from the PMCF (see Figure 4). Furthermore, the quality criteria and metrics were obtained from existing literature in a related review as well as from conducted interviews with domain experts from the field of BPM. 

\textbf{Literature review:} In literature, there exists numerous works focusing on process model quality in order to, on one the hand, improve the creation and, on the other hand, to foster the comprehension of process models \cite{figl2017comprehension,dikici2018factors}. Moreover, different frameworks as well as guidelines with an emphasis on process model quality were defined for this context as well (see Section 1). Regarding the PMCF, several data sources and publication libraries (e.g., Google Scholar, SpringerLink, IEEE Xplore Digital Library) were examined for works concerning process model quality during process model creation and comprehension. Therefore, search strings were elaborated with the combinations of keywords derived from our knowledge of the subject focus (e.g., process model quality).

\textbf{Interviews:} In the conducted interviews, eight domain experts (i.e., from academia and industry) with many years of expertise in the field of BPM were personally consulted. The used catalog of questions contained a set of opinion (e.g., Which quality aspects in a process model are essential in the creation of a complete and correct process documentation, but also to ensure that the created model can be comprehended by all involved stakeholders?), behavioral (e.g., How can it be avoided that a process modeler creates an incorrect model?), and competency (e.g., When does the application of a process modeling language or the creation/comprehension of a process model become too complex?) questions in terms of preservation of quality in process modeling as well as process model comprehension. From the given answers of the interviewed domain experts, quality criteria and metrics were identified referring on the perspective of process modelers, readers, and synergies covering both perspectives. 

The categorization of the quality criteria and corresponding metrics were done within a consensus decision-making process, involving participants from industry and academia as well as obtained insights from the literature review, including the interviews. 

As a result, the perspective of process modeler comprises the following six main quality criteria:

\begin{itemize}
	\item Process Modeling Language: This criterion covers crucial aspects (e.g., workflow patterns) that a process model language should support for the creation of high quality process models.
	\item Process Modeling Tool: The quality of created process models are dependent of the used process modeling tool. Hence, this criterion summarizes vital aspects (e.g., process views) about tool-support in process modeling.
	\item Information: This criterion is concerned with process information retrieval and addresses aspects like correctness and completeness of process information.
	\item Errors: Semantic (e.g., logical errors) and syntactic (e.g., errors in modeling conventions) errors in a process model are subject of this criterion.
	\item Person: Person-related characteristics (e.g., process modeling experience) are considered in this criterion.
	\item Process Modeling Guidelines: This criterion covers guidelines and rules (i.e., from enterprise and academic) that were defined in order to create process models of high quality.
\end{itemize} 

These quality criteria are subdivided into several sub-metrics, resulting in 54 different quality metrics. 

Regarding the perspective of process reader, 42 quality metrics are summarized in a total of seven main quality criteria, which are defined as follows:

\begin{itemize}
	\item Process Modeling Language: This criterion addresses aspects (e.g., modeling language complexity) that define comprehensible process modeling language.
	\item Medium: The subject of this criterion is the question with which medium (e.g., paper-based) is the process model comprehended.
	\item Information: This criterion deals about which kind of process information (e.g., process participants) are included in the process model.
	\item Person: Person-related characteristics (e.g., process modeling experience) are considered in this criterion.
	\item Level of Detail: In this criterion, the level of detail (e.g., abstract or concrete) of the comprehended process model is addressed.
	\item Representation Factors: Aspects about the process model representation (e.g., number of elements) and the model structure (e.g., block structure) are subject of this criterion.
	\item Comprehension Questions: Process model comprehension performance analysis (e.g., comprehension questions) are considered in this criterion.
\end{itemize} 

Altogether, the ETT contains 96 quality metrics. Each of the 96 metrics can be assigned to one or more quality types in the PMCF (e.g., learnability of a modeling language refers to K1 in Figure 4). Further, the evaluation of the quality types facilitates the identification of noise (e.g., difference in process model knowledge) between process modelers and readers as known from the PMCF. Due to space limitations, Figure 5 only presents an excerpt of the ETT (see Appendix A).

\begin{figure}[h]
	\centering
	\includegraphics[width=\linewidth]{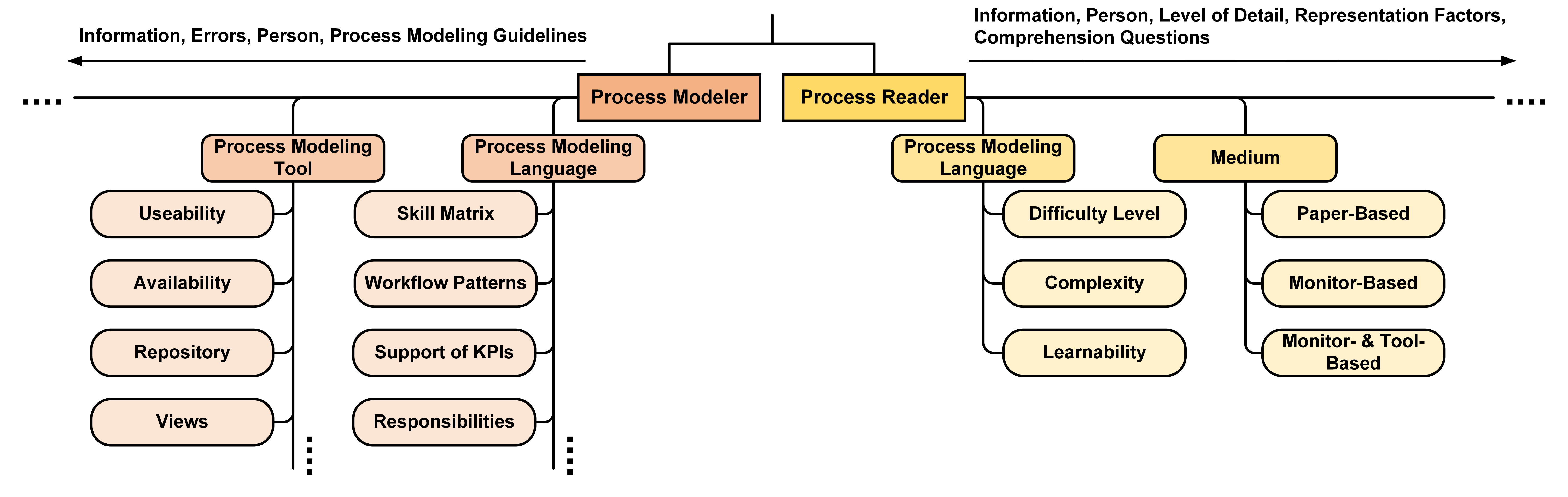}
	\caption{Excerpt from the ETT}
	\label{fig:ettt}
\end{figure}

In order to measure and quantify process model comprehension, the importance and the impact on process model comprehension for each quality criterion and its related metric in the ETT had to be determined. For example, as shown in \cite{article}, the number of elements in a process model constitutes a more critical factor having a stronger impact on process model comprehension juxtaposed to the labeling of process model elements. For this reason, a survey with 131 participants from academia as well as industry was conducted. The participants of the survey were asked to rate and place the quality criteria as well as related metrics for both considered perspectives (i.e., process modeler and reader) in an order from important to unimportant to determine the rank and the impact of each quality criterion and metric. For the determination of the rank, the results were analyzed with the weighted arithmetic mean $ \bar{X} $ that is defined as follows:


\begin{equation}
	\bar{X} = \frac{ \sum_{k = 1}^{n}{ w_{k} }{p_{i,k}}}{ \sum_{k = 1}^{n}{ w_{ k } } }   , 
\end{equation}

where $ k $ is the rank, $n$ is the number of ranks,  $ w_{k} $ is the weighting factor for the rank k, $ i $ is the quality metric, and $ p_{i,k} $ is the percentage choice of the quality metric $ i $ on the rank $ k $. 

The weighting factor $ w_{ k } $ had to be determined, since each quality criterion and related metric exert a different impact on process model comprehension. For the calculation of the weighting factor $ w_{ k }  $, the following evaluation methods for information retrieval were juxtaposed in a series of repeated measurements: Rank Sum, Reciprocal Rank, Rank Exponent, Discounted Cumulative Gain, and Distance Normalized Logarithm \cite{sakai2007reliability}. Most considered methods showed a disparate differentiation between the quality criteria and metrics (i.e., Rank Sum, Rank Exponent, Discounted Cumulative Gain). Moreover, the methods presented no standardized distance between the highest and the lowest rank as well the ranks in between. In more detail, the provided weighting factors of those methods exhibited a limited growth leading to an incorrect differentiation of the quality criteria and metrics. However, further analyses demonstrated that a cubic (i.e., Reciprocal Rank) or an exponential growth (i.e., Distance Normalized Logarithm (DNLog)) should be considered in this context. A comparison of both methods revealed that the DNLog was more suitable for our purpose. The reason for this decision was that the DNLog ensures a more detailed weighting of the quality criteria as well as metrics. For example, a vital criterion or metric (e.g., syntactical process model correctness \cite{leopold2015learning}) has a more significant impact on process model comprehension and should be given a greater weight compared to negligible ones (e.g., avoid OR routing process model elements \cite{mendling2010seven}). As a result, the DNLog was chosen for the calculation of the weighting factor $ w_{ k } $, which is defined as:

\begin{equation}
	w_{ k } = 10^{(n - k)  \frac{\log_{10}(d)}{(n - 1)}}   ,
\end{equation}

where $ n $ is the number of items, $ k $ is the rank, and $ d $ is the score in the survey. 

Equations (1) and (2) were used to analyze the responses from the 131 participants of the survey and to determine the rank as well as the impact of all quality criteria and related metrics on process model comprehension for the perspective of process modelers, readers, and the interaction between both perspectives. 

\section{Implementation of the Process Model Comprehension Framework}

This section presents how the PMCF is implemented in order to measure and quantify process model comprehension. For this purpose, the ETT has been implemented in a Microsoft Excel workbook template (see Appendix B). According to the Communication Theory, this workbook is used to evaluate process models regarding their comprehensibility to unravel noise between the communication of process modelers and readers. Note that the workbook consists of nine sheets:

\begin{center}
	\begin{tabular}{ll}
		\raisebox{.5pt}{\textcircled{\raisebox{-.9pt} {1}}} Configuration		&	\raisebox{.5pt}{\textcircled{\raisebox{-.9pt} {2}}} Process Modeling Language Complexity  \\
		\raisebox{.5pt}{\textcircled{\raisebox{-.9pt} {3}}} Supported Patterns		&  	\raisebox{.5pt}{\textcircled{\raisebox{-.9pt} {4}}} Quality Metrics\\
		\raisebox{.5pt}{\textcircled{\raisebox{-.9pt} {5}}} Questionnaire for Process Modeler		&  	\raisebox{.5pt}{\textcircled{\raisebox{-.9pt} {6}}} Questionnaire for Process Reader\\
		\raisebox{.5pt}{\textcircled{\raisebox{-.9pt} {7}}} Perspective of Process Modeler	& 	\raisebox{.5pt}{\textcircled{\raisebox{-.9pt} {8}}} Perspective of Process Reader \\
		\raisebox{.5pt}{\textcircled{\raisebox{-.9pt} {9}}} Summary		& 
	\end{tabular}
\end{center}

The workbook supports the evaluation of process models expressed in terms of the following process modeling languages: Business Process Modeling and Notation (BPMN) 2.0, Event-driven Process Chains (EPCs), and UML Activity Diagrams. Since the relevant sheets are predefined with the results obtained from the survey (i.e., weighting of the quality criteria and metrics), only the sheets (4), (5), and (6) must be completed to quantify the perspectives (i.e., process modeler, reader, and both) in the context of process model comprehension. Changes in the remaining sheets are only necessary if factors (e.g., weighting factor $w_k$) shall be adjusted, further quality criteria and metrics need to be introduced, or the workbook shall be extended to support additional modeling languages. Regarding the latter, in all sheets, respective information for the newly introduced modeling language must be added. 

\textbf{(1) Configuration:} Process modeling languages have various impact on process model comprehension (e.g., in terms of expressiveness) \cite{zimoch2017eye}. Thus, the operationalization thereof is performed in the workbook in (2). Moreover, with respect to comparability, all results in the workbook are normalized within an interval between $[1,10]$. Thereby, a 1 represents a worse outcome, whereas a 10 indicates the best outcome regarding process model comprehension. Hence, in this sheet, the complexity coefficient $ \|C_i \| $ calculated in (2) (see Equation (5)) is normalized to $ \bar{C} _{i} $ for each process modeling language within an interval between $[1, 10] $, and is defined as follows:

\begin{equation}
	\bar{C}_i = 10 - {\frac{(10 * \|C_i \|)-\|C_i \|}{MAX(C_n) * 10}},
\end{equation}

where $i$ is the process modeling language, $\|C_i \|$ is the complexity score for the specific process modeling language $ i $, and $ C_{n} $ is the set of all complexity scores.

Another factor having an impact on process model comprehension is the use of workflow patterns in a process model. These patterns play a crucial role in the creation of such models and are, therefore, especially for process modelers of importance \cite{white2004process}. The impact of workflow patterns, (i.e., $P_i$ calculated in (3), currently just for control flow patterns) is shown in this sheet and is included in the determination of the process model comprehension score for the perspective of process modelers (see (7)).

\textbf{(2) Process Modeling Language Complexity:} In general, a process modeling language is composed of a number of modeling elements, their characteristics (e.g., different activity types), and their relations (e.g., different flow types such as sequence or data), which define the expressiveness of respective language \cite{list2006evaluation}. Based on this consideration, the workbook defines the complexity of a process modeling language $C_{i}$ as a three-dimensional vector:

\begin{equation}
	C_i = (x_i, y_i, z_i),
\end{equation}  

where $i$ is the process modeling language, $ x_i $ is the number of elements of the modeling language, $y_i$ is the number of characteristics per element, and $ z_i $ is the number of relationships per element \cite{LaueG06}. 

Accordingly, the number of elements, their characteristics as well as their relations reflect the complexity of a modeling language. With the Euclidean norm, $C_i$ can be converted to the complexity score $\|C_i \|$ for a specific process modeling language $i$:

\begin{equation}
	\|C_i \|  = \sqrt{x_i^2 + y_i^2 + z_i^2}
\end{equation}

\textbf{(3) Supported Patterns:} The expressiveness as well as suitability of a process modeling language is not only determined by the number of elements, their characteristics, and their relations (see (2)), but also by the number of supported workflow patterns \cite{white2004process}. Workflow patterns describe specific mechanisms supporting stakeholders dealing with the complexity of process models (e.g., consideration of different perspectives such as control flow and data). For this reason, \cite{van2009workflow} defined a set of workflow patterns, which are considered in the workbook. In this context, the workbook supports the following workflow pattern types: control flow, data, and resource patterns. Furthermore, for each workflow pattern type, the workbook considers which workflow patterns are fully, partially, or not supported in the respective modeling language. Based on this consideration, the score for supported patterns $P_i$ is determined as follows:

\begin{equation}
	P_i = m_i + n_i + o_i,
\end{equation}

where $i $ is the process modeling language, $m_i$ is the number of fully and partially supported control flow patterns, $n_i$ is the number of fully and partially supported data patterns, and $o_i$ is the number of fully and partially supported resource patterns.

$P_i$ (in percentage, currently just for control flows patterns) is used in (1) for the determination of the process model comprehension score pertaining to the perspective of process modelers (see (7)).

\textbf{(4) Quality Metrics:} In this sheet, for the perspective of process modelers and readers, the quality metrics from the PMCF (i.e., ETT) related to the evaluated process model are determined. In particular, for each quality metric, an explanation of the respective metric is given as well as an instruction on how to determine the corresponding metric (e.g., number of in-/outgoing edges per process modeling element). The metrics are determined either as described in respective literature or must be determined manually considering the process model to be evaluated. If the determined result has not yet been normalized, the result will be normalized in an additional step within an interval between $[1, 10]$. Thereby, a result towards the right boundary (i.e., 10) describes a more positive impact on the comprehension of the process model.

\textbf{(5) Questionnaire for Process Modeler:} The comprehension of a process model depends not only on factors of the respective process model (e.g., size of the process model), but also on the perception  of the original creator (i.e., process modeler) of the model. Thereby, a process modeler has personal related characteristics (e.g., expertise in process modeling) in the context of process modeling as well as an individual interpretation about the information and knowledge regarding the process and its related model. Furthermore, there exist a specific mental interpretation about the process and resulting process model in the mind of the process modeler. As described in Section 2.1, noise may occur in the communication of process information and knowledge between the process modeler as well as readers. For this reason, it is important to capture and know both personal related characteristics as well as the interpretation of the process modeler (i.e., perspective of process modelers) to identify respective noise and to initiate  countersteps. Therefore, the original process modeler of a corresponding model has to answer a specific questionnaire capturing personal related characteristics as well as the related interpretation of the process and its resulting process model. The questionnaire consists of a set of 49 different questions addressing quality criteria and metrics from the ETT to capture the perspective of the process modeler. The questions types are a set of true-or-false and Likert scale questions. The responses are compiled to a score within the interval $[1, 10]$, whereas ten indicates a more positive impact on process model comprehension.

\textbf{(6) Questionnaire for Process Reader:} Similar to the process modeler, a specific questionnaire to capture the perspective of process readers has to be answered. This questionnaire consists of 24 questions related to corresponding quality criteria as well as metrics from the ETT in order to gather the perception and interpretation of process readers about the comprehended process model. Equally to (5), the responses reflect a score within the interval $[1, 10]$, whereas ten constitutes the best score regarding process model comprehension.

\textbf{(7) Perspective of Process Modeler:} This sheet contains all the ranked as well as weighted quality criteria and corresponding metrics from the ETT for process modelers. The individual scores are automatically generated based on the results obtained from the sheets (1), (4), and (5). Hence, this sheet requires no interaction and is used exclusively for aggregating and calculating the process model comprehension score for process modelers. Therefore, as a first step, the sum of all quality metrics $Q_c$ of the respective criterion is calculated:

\begin{equation}
	Q_c = \sum_{i = 1}^{n}i,
\end{equation}

where $c$ is the quality criterion, $i$ is the quality metric, and $n$ is the number of quality metrics.

The final process model comprehension score for process modelers $S_m$, which represents a score within the interval $[1, 10]$ (i.e., ten is the best), is built from the sum of all aggregated quality criterion $Q_c$:

\begin{equation}
	S_m = \sum_{c = 1}^{6} = Q_c
\end{equation}

Based on $Q_c$, possible factors for noise can be identified by considering related quality metrics with a score towards the left boundary within the interval $[1, 10]$ (see Section 5).

\textbf{(8) Perspective of Process Reader:} Similar to (7), the process model comprehension score for process readers $S_r$ is determined in this sheet. Hence, all ranked as well as weighted quality criteria and corresponding metrics from the ETT are shown here. The determination of the score is carried out in the same way as described in (7), only with relevant aspects for process readers. Therefore, no changes are required in this sheet. The process model comprehension score for process readers reflects a score within the interval $[1, 10]$ (i.e., ten is the best), based on the sum of the aggregated quality criteria for process readers. As with the process modeler, factors for noise in the comprehension of a process model can be identified on the basis of the individual calculated scores for respective quality criterion and related metrics (see Section 5).

\textbf{(9) Summary:} The final sheet in the workbook presents the quantified process model comprehension scores on the evaluated process model. Here, the scores for the perspective of process modelers $ S_m $ and readers $ S_r $ as well as the interaction of both $S_b$ are presented. The single scores are within the interval $[1, 10]$, whereas 1 indicates the worst score regarding process model comprehension and 10 the best. Thereby, the scores for process modelers and readers are determined in the sheets (7) and (8), respectively. The score for the interaction of both perspectives $S_b$ is determined as follows:

\begin{equation}
	S_b = (w_m * S_m) + (w_r * S_r),
\end{equation}

where $w_m$ is the weight for process modelers and $w_r$ is the weight for process readers.

The two weights $w_m$ and $w_r$ were determined within the survey with the specific question asking about which aspect in a process model is considered to be more significant, i.e., ease of creation ($w_m$) or ensuring proper comprehensibility ($w_r$). Hence, the percentage distribution was calculated from the responses given. 

\section{Case Study and Application of the Process Model Comprehension Framework}

In order to demonstrate the applicability of the PMCF (i.e., workbook), a case study with 33 participants from industry was conducted (see Appendix C). According to collected demographic data, all participants stated that they already had been working with process models. Hence, the experience in process modeling as well as model comprehension was between a novice and intermediate level. The participants were asked to comprehend five real-world scenarios from a business consultant company. Each scenario was documented in two different process model variants (i.e., ten in total), with a respective emphasis on the following process modeling aspects: start events, end events, loops, parallelism, and decomposition. In one variant, mentioned aspects were explicitly documented in a process model, while in the other models, they were only implicitly (i.e., described in an activity) documented. In addition, for each process model, participants needed to answer a set of four true-or-false comprehension questions about semantic aspects in the models. Regarding the PMCF, the workbook sheets (4) Quality Metrics, (5) Questionnaire for Process Modeler, and (6) Questionnaire for Process Reader were completed accordingly. Thereby, the sheet (4) was completed by considering the characteristics of the process models (e.g., number of modeling elements). The sheet (5) was answered by the original creator of the process models. Thereby, there was only one original creator for each process model. Finally, the sheet (6) was answered by the 33 participants of the study after each comprehended process model. Table 1 presents the results from the case study. In detail, the table shows, for each process model and respective variant, the mean of the results from the comprehension questions (i.e., max is four) as well as the determined process model comprehension scores with the workbook for the process modeler, reader (i.e., average), and both (i.e., average). 

\begin{table}[h]
	\centering
	\caption{Demonstration of the applicability of the PMCF}
	\begin{tabular}{l|c|ll|l|c|ll|}
		\cline{2-4} \cline{6-8}
		& \multicolumn{3}{c|}{\textbf{Variant 1 (Explicit)}}                                &  & \multicolumn{3}{c|}{\textbf{Variant 2 (Implicit)}}                                \\ \cline{1-4} \cline{6-8} 
		\multicolumn{1}{|l|}{\textbf{Process Model}}                    & \textbf{Result} & \multicolumn{2}{c|}{\textbf{Perspective}} &  & \multicolumn{1}{c|}{\textbf{Result}} & \multicolumn{2}{c|}{\textbf{Perspective}} \\ \cline{1-4} \cline{6-8} 
		\multicolumn{1}{|l|}{\multirow{3}{*}{\begin{tabular}[c]{@{}l@{}}Process Model 1\\ (Start Event)\end{tabular}}} & \multirow{3}{*}{2.76}     & Modeler               & 5.20          &  & \multirow{3}{*}{1.56}     & Modeler              & 5.19           \\ \cline{3-4} \cline{7-8} 
		\multicolumn{1}{|l|}{}                      &                            & Reader            & 6.35        &  &                            & Reader            & 6.30          \\ \cline{3-4} \cline{7-8} 
		\multicolumn{1}{|l|}{}                      &                            & Both             & 6.17          &  &                            & Both            & 6.14            \\ \cline{1-4} \cline{6-8} 
		\multicolumn{1}{|l|}{\multirow{3}{*}{\begin{tabular}[c]{@{}l@{}}Process Model 2\\ (End Event)\end{tabular}}} & \multirow{3}{*}{2.38}     & Modeler               & 5.39          &  & \multirow{3}{*}{2.00}     & Modeler              & 5.39            \\ \cline{3-4} \cline{7-8} 
		\multicolumn{1}{|l|}{}                      &                            & Reader            & 6.37         &  &                            & Reader            & 6.39            \\ \cline{3-4} \cline{7-8} 
		\multicolumn{1}{|l|}{}                      &                            & Both             & 6.22           &  &                            & Both            & 6.23          \\ \cline{1-4} \cline{6-8} 
		\multicolumn{1}{|l|}{\multirow{3}{*}{\begin{tabular}[c]{@{}l@{}}Process Model 3\\ (Loop)\end{tabular}}} & \multirow{3}{*}{1.82}     & Modeler               & 4.74           &  & \multirow{3}{*}{1.06}     & Modeler              & 4.70            \\ \cline{3-4} \cline{7-8} 
		\multicolumn{1}{|l|}{}                      &                            & Reader            & 6.29         &  &                            & Reader            & 6.26            \\ \cline{3-4} \cline{7-8} 
		\multicolumn{1}{|l|}{}                      &                            & Both             & 6.06           &  &                            & Both            & 6.04            \\ \cline{1-4} \cline{6-8} 
		\multicolumn{1}{|l|}{\multirow{3}{*}{\begin{tabular}[c]{@{}l@{}}Process Model 4\\ (Parallelism)\end{tabular}}} & \multirow{3}{*}{1.94}     & Modeler               & 4.69           &  & \multirow{3}{*}{1.65}     & Modeler              & 4.70          \\ \cline{3-4} \cline{7-8} 
		\multicolumn{1}{|l|}{}                      &                            & Reader            & 6.47           &  &                            & Reader            & 6.28          \\ \cline{3-4} \cline{7-8} 
		\multicolumn{1}{|l|}{}                      &                            & Both             & 6.20           &  &                            & Both            & 6.05           \\ \cline{1-4} \cline{6-8} 
		\multicolumn{1}{|l|}{\multirow{3}{*}{\begin{tabular}[c]{@{}l@{}}Process Model 5\\ (Decomposition)\end{tabular}}} & \multirow{3}{*}{2.47}     & Modeler               & 5.90           &  & \multirow{3}{*}{2.19}     & Modeler              & 5.89          \\ \cline{3-4} \cline{7-8} 
		\multicolumn{1}{|l|}{}                      &                            & Reader            & 6.38          &  &                            & Reader            & 6.35          \\ \cline{3-4} \cline{7-8} 
		\multicolumn{1}{|l|}{}                      &                            & Both             & 6.30          &  &                            & Both            & 6.29           \\ \cline{1-4} \cline{6-8} 
		\multicolumn{8}{l}{\textbf{Note:} Perspective scores range within the interval $[1, 10]$, whereas ten} \\
		\multicolumn{8}{l}{indicates the best score regarding process model comprehension} 	
	\end{tabular}
\end{table}

According to the results from the comprehension questions, the variants with explicitly documented process modeling aspects had a more positive impact on process model comprehension (i.e., higher comprehension scores). Considering both perspectives, in general, the process model comprehension scores mainly confirm this observation (i.e., higher perspective scores). Only for the second implicit process model (i.e., end event) variant, the score is slightly higher. Consider Table 1, there are only small differences in the comprehension scores between the perspectives, compared to the differences in the comprehension questions. A reason is that the use of questions represents a simple metric, which is susceptible to deviations (e.g., guessing or heterogeneous distribution of expertise). The PMCF, in turn, not only considers the performance in model comprehension (i.e., answering of the questions), but also a variety of quality metrics, which each have a different strong impact on process model comprehension, leading to a more fine-grained result. Furthermore, which is not apparent from the consideration of the comprehension question results only, there are differences between the process modelers and readers. Regarding the process readers, the PMCF workbook results in a comprehension score of about 6. Since the comprehension score is within the interval $[1, 10]$ (i.e., 10 is the best), it indicates that the process models are slightly above the average in terms of process model comprehension. Furthermore, it is remarkable that the original creator of the process models evaluated their own created process models as less comprehensible in the retrospect compared to respective readers. The comprehension scores for process modelers are approximately between 4 and 6. A reason could be that the process modelers during the answering of the PMCF worksheet (5) Questionnaire for Process Modeler have critically recapitulated their own process model. More specifically, single items from the PMCF worksheet (5) (e.g., knowledge about process domain, correctness of process information) may had drawn attention to possible deficits in the process model. Since process modelers and readers have different perspectives, a uniform comprehension of the process models used in the study was not given, due to occurring noise in the communication of process information and knowledge. 

\subsection{Application of the Process Model Comprehension Framework}

The PMCF allows for the identification of reasons for difficulties as well as noise (e.g., discrepancies in process domain knowledge) during the comprehension of the presented process models in order to initiate steps to improve respective models. Therefore, the workbook sheets (7) Perspective of Process Modeler and (8) Perspective of Process Reader may be considered. As described in Section 4, these sheets are aggregating and calculating the process model comprehension scores for respective perspectives. For this purpose, the sum of all  quality metrics for each quality criterion are calculated (i.e., six for process modeler and seven for process reader; see Section 4). Afterwards, the final comprehension score is determined from the sum of the aggregated quality criteria and compiled to a score within the interval $[1, 10]$, whereas 10 indicates the best score regarding process model comprehension. In the optimum case, the final comprehension score is 10, which means that the quality criteria and metrics have also been aggregated to a value of 10. \\
In the following, an example is presented how the insights from the PMCF worksheets may be used in order to foster process model comprehension. Therefore, the results regarding the third process model (i.e., explicit loop) from the process modeler and a reader are considered (see Appendix D).

\textbf{Perspective of Process Modeler:} Considering the individual scores we obtained from the case study, for example, we noticed for the perspective of the process modeler that the score regarding the quality criterion Information (see Section 3) is 5.01. Thereby, the quality criterion Information is concerned with process information retrieval and consists of the following metrics: completeness (i.e., Is the process information complete?), correctness (i.e., Is the process information correct?), availability (i.e., What availability does the process information have?), and method (i.e., Which methods are available for process information retrieval?). Regarding the two latter metrics, in our example, the score is 2.1 for availability and 1.6 for method. As a direct consequence, the original process modeler had difficulties with the availability of process information (i.e., only textual process documentations were available) as well as in the choice of methods for process information retrieval (i.e., only the study of the textual process information was available). Therefore, an increase in the availability of process information and methods for information retrieval would, on the one hand, lead to the creation of a better comprehensible process model because process information can be collected more effectively. On the other hand, as a result, an increase in these two metrics would have a positive effect on the score regarding the quality criterion Information, thus leading to an increase in the final process model comprehension score for the process modeler, reader and the interaction between both perspectives.

\textbf{Perspective of Process Reader:} Considering the perspective of process readers and their individual scores obtained from the case study, for example, the score regarding the quality criterion Person (see Section 3) is 5.34 in our example. In more detail, the process model readers have stated that their experience of working with process models (e.g., number of analyzed process models) is maximum at the level of an intermediate. Moreover, contemplating the quality criterion Representation Factors and related metrics that are concerned with structural factors of the process models (e.g., block structure), the score is 3.68. These scores provide us with indications about how to increase the final comprehension score for the perspective of process reader. On the one hand, process readers should be more concerned with different kind of process models in order to increase their experience in working with such models. Further, on the other, the used process models in the study could be adjusted by respecting a consistent block structure. These steps would then have a positive impact on the final comprehension score of process readers but also on the comprehension score of process modeler as well as the interaction between both.

In summary, the conducted case study demonstrated the successful application of the PMCF in a practical environment. The results indicated that there is a non-uniform comprehension of process models between process modelers and readers. Moreover, the PMCF revealed that process modelers and readers are confronted with different challenges (i.e., noise) in process model comprehension. In general, although the PMCF is still in the early stage of development, it can already be applied on real-world process models of organizations for identifying potentials for process model improvements, i.e., in order to prevent noise in the communication of process information and knowledge, thus fostering the general comprehension of such models.

\subsection{Limitations}

Although the PMCF is still applicable in practical environments, it is noteworthy that it is still in an early stage of development and that the PMCF is a first step towards measuring and quantifying process model comprehension. Hence, the PMCF as well as the developed workbook are currently confronted with limitations that need to be discussed and will be subject of future work.  First, the interpretation of the calculated process model comprehension scores must be evaluated. In detail, the results range in the interval between $[1, 10]$, whereas 10 indicates the best score regarding process model comprehension. The first applications of the workbook demonstrated that the calculated scores are reliable (see Section 5). However, the workbook must be applied on many more process models in order to be able to interpret differences in the scores accurately and to define a score threshold, from which process models are well comprehensible for the general public. Second, the use of the DNLog in order to determine the weights (i.e., $w_k$) for the quality criteria as well as metrics should be further evaluated. Third, the normalization of the results from the quality criteria and related metrics as well as the meaningful applicability of the normalization approach need to be further evaluated as well. Fourth, in general, the completeness, accuracy, and validity of the PMCF as well as the workbook need to be scrutinized in detail as the PMCF contains a great number of quality criteria and metrics. These quality aspects are derived from the process of adaption of the CMQF, a literature review, and conducted interviews. However, myriads of quality aspects exist that currently do not fall within the scope of the PMCF \cite{dikici2018factors,borthick2016detecting}. These include, in particular, cognitive aspects, which, as known from recent studies in this context, exert a significant impact on the comprehension of process models \cite{haisjackl2018humans,zimoch2018utilizing}. Moreover, cognitive aspects may be the critical mediator in the communication of process information as well as knowledge between process modelers and readers. 

\subsection{Implications} 

With this paper, we highlight the important implications of the PMCF and the ability to measure and to quantify process model comprehension for practice. Process models constitute vital artifacts in the application of information technologies (e.g., PAIS). In particular, during the utilization of information systems, undiscovered errors made (e.g., incorrect process documentation due to noise in the communication of process information and knowledge) may have critical impacts in the later utilization and, hence, projects might not deliver the required results or even fail. For this reason, it is of importance that process models are created correctly as well as accurately. At the same time, it should be ensured that these models are comprehensible for all involved stakeholders. In this context, a process model is an artifact used for the communication of process information and knowledge between participants (see Section 2.1). During this kind of communication, noise may occur that may impairs the comprehension of process models. Therefore, during process model comprehension, the PMCF allows for the measurement and quantification of the perspectives of process modelers, readers, and the interaction between both. This allows for the identification of noise, which could therefore be addressed ensuring a proper process model comprehension. Moreover, since the PMCF covers different quality criteria and metrics covering various aspects (e.g., process modeling tools, medium (see Section 3.1)) for respective perspectives, organizations are able to identify concrete deficiencies in the context of process models with the provided scores of the PMCF. Further, despite the preliminary focus on the comprehension of process models, the insights obtained with the PMCF also affect the creation of process models (see Section 5.4). Thus, in the creation or optimization of process models, factors for noise in the communication of process information and knowledge can be avoided paving the way for process models of high quality. With the support and extensibility of additional process modeling languages, the PMCF assists organizations in the selection of an appropriate modeling language. This is applicable when a process modeling language is selected for the first time (e.g., in the early phases of the information systems development process), or in case of a modeling language change, which is, for example, pursued due to a process model redesign.  

\subsection{Future Work} 

The used approach of definition for the PMCF allows for an appropriate extensibility. More specifically, novel quality criteria or metrics can be added to the ETT. Therefore, to address the discussed limitations (see Section 5.2), the PMCF is currently used in ongoing studies that evaluate different process models from theory as well as practice with heterogeneous participant groups. The objective is, on the one hand, to improve our general interpretation of the calculated process model comprehension scores and, on the other hand, to identify additional noise factors in the communication of process knowledge and information between process modeler as well as reader. The unraveled insights allow for the definition of directives towards creating better comprehensible process models with high quality. In this context, the results for the different perspectives obtained from the PMCF are juxtaposed with existing rules and guidelines (e.g., Guidelines of Modeling \cite{becker2000guidelines}, Seven Process Modeling Guidelines \cite{mendling2010seven}, which are intended to ensure a proper comprehension of process models, in order to evaluate their contribution regarding process model comprehension. Moreover, the weighting factor $ w_k $ is examined in detail and will be adjusted as well as refined accordingly when, for example, new quality criteria and metrics are added. In addition, other approaches, in addition to the already evaluated one (see Section 3) to determine the weighting factor $ w_k $ are juxtaposed to the DNLog in order to evaluate their appropriateness. Furthermore, the PMCF will be extended and enriched with further quality criteria and metrics to obtain more fine-grained scores. In this context, we are currently augmenting the PMCF with additional criteria and metrics to include the creation of process models (i.e., process of process modeling) \cite{burattin2019learning}. This augmentation should ensure that process models are created in a high quality and in a comprehensible form from the very beginning and, thus, should prevent the occurrence of noise in the communication of process knowledge as well as information. Support for additional process modeling languages and workflow patterns are subject of future work. Finally, to pave the way for cognitive aspects, the PMCF will be integrated into the conceptual framework of the authors that incorporates concepts from cognitive neuroscience and psychology introduced in \cite{ZimochPPSR17}. 

\section{Conclusion}

This paper presented the Process Model Comprehension Framework (PMCF) as a first step towards measuring and quantifying the comprehensibility of process models. Based on the Communication Theory and the CMQF, the PMCF considers the perspectives of process modelers and readers as well as the interaction between them as main determinants in process model comprehension. Therefore, in order to identify and prevent noise (i.e., misinterpretation in the communication of process information and knowledge due to person- and model-related characteristics) in model comprehension, an ETT was defined composed of a set of quality criteria and 96 metrics in total. Thereby, the quality criteria and related metrics were obtained from interviews with experts in the field of Business Process Management and in a literature review. Further, these quality aspects are ranked and weighted with regard to their importance and impact on process model comprehension in a survey with 131 participants from academia and industry. The ETT and the results from the survey have been implemented in an Excel workbook, which allows to measure and quantify process model comprehension on existing process models. The application of the workbook and how to improve process models based on the results obtained was demonstrated successfully in a case study with 33 participants from industry. Accordingly, the PMCF and its corresponding workbook shall contribute to identify and avoid pitfalls (i.e., noise) in the communication of process knowledge and information between process modelers and readers. In addition, the PMCF wants to ensure that process models implemented in information systems are of high quality for the purpose of a proper model comprehensibility. Therefore, the calculated process model comprehension scores with the PMCF and related workbook serve as a signpost in order to foster and ensure a correct comprehension of process models. Further, the initial creation of comprehensible process models or the optimization of existing models is supported by the PMCF. In general, the PMCF and the future work thereof shall assist organizations in all phases (i.e., design, implementation, and management) in the utilization of information systems.

\appendix
\section*{Appendix}

\section{}
The complete depiction of the ETT can be found at: 
\url{https://tinyurl.com/wtfmh9q}
\section{}
The workbook template can be found at: 
\url{https://tinyurl.com/roky2tf}
\section{}
Study materials can be found at: 
\url{https://tinyurl.com/yx3ht8ry}
\section{}
The example worksheet can be found at: 
\url{https://tinyurl.com/y8p3yjcs}

\bibliographystyle{unsrt}  
\bibliography{ArXiv_Winter}

\end{document}